\documentclass{jfm}

\pdfoutput=1

\usepackage{graphicx}
\usepackage{bm}
\usepackage{amssymb, amsmath}
\usepackage{url}
\usepackage{color}
\usepackage{soul}
\definecolor{orange}{rgb}{1,0.5,0}

\begin{document}

\shorttitle{Robust wall states in RRBC} 
\shortauthor{Favier and Knobloch} 

\title{Robust wall states in \\rapidly rotating Rayleigh-B\'enard convection}

\author
 {
 Benjamin Favier\aff{1}
  \corresp{\email{favier@irphe.univ-mrs.fr}}
  and
 Edgar Knobloch\aff{2}
  }

\affiliation
{
\aff{1}
Aix Marseille Univ, CNRS, Centrale Marseille, IRPHE, Marseille, France
\aff{2}
Department of Physics, University of California, Berkeley, CA 94720, USA
}

\maketitle

\begin{abstract}
We show, using direct numerical simulations with experimentally realizable boundary conditions, that wall modes in Rayleigh-B\'enard convection in a rapidly rotating cylinder persist even very far from their linear onset. These nonlinear wall states survive in the presence of turbulence in the bulk and are robust with respect to changes in the shape of the boundary of the container. In this sense, these states behave much like the topologically protected states present in two-dimensional chiral systems even though rotating convection is a three-dimensional nonlinear driven dissipative system. We suggest that the robustness of this nonlinear state may provide an explanation for the strong zonal flows observed recently in experiments and simulations of rapidly rotating convection at high Rayleigh number.
\end{abstract}

\section{Introduction}
In recent years there has been considerable interest in the so-called topologically protected states on account of their robustness with respect to system perturbations. These states, originally discovered in the context of two-dimensional semiconductors, take the form of a unidirectional current along the boundary. This current persists under system perturbations, including defects and changes in the boundary, its persistence guaranteed by topological arguments. Other examples include two-dimensional chiral materials \citep{gollub,irvine,irvine1}, isostatic lattices in two dimensions \citep{kane} and photonic systems \citep{khanikaev}. Recently it was observed that similar arguments apply to equatorial shallow-water waves and used to confirm the presence of two types of low-frequency eastward-propagating equatorially trapped waves, Kelvin and Yanai waves, separating bulk waves at higher positive and negative frequencies \citep{marston,venaille}. The theory predicts that both these boundary currents are robust, for example, with respect to topography. This type of theory applies to linear dissipationless systems exemplified by the Schr\"odinger equation in quantum systems \citep{thouless} and the related shallow-water system studied by \cite{marston}. Odd viscosity can be added without changing the conclusions of the theory because this type of viscosity does not dissipate energy \citep{banerjee,souslov,venaille1}.

\begin{figure}
\begin{center}
\includegraphics[width=0.9\textwidth]{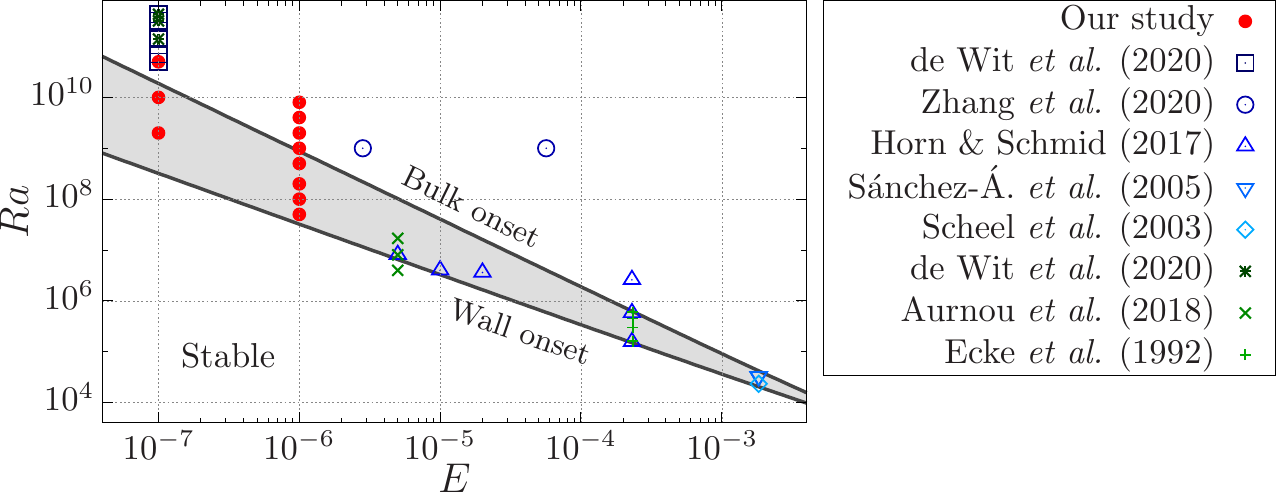}
\caption{Summary of the runs performed for this study (full symbols \textcolor{red}{$\bullet$}), along with past numerical \citep{scheel2003,sanchez2005square,horn2017prograde,kunnen,bodenschatz} and experimental \citep{zhongEK,aurnou2018rotating,kunnen} studies. The lower black line corresponds to the onset of wall modes as predicted by linear theory while the upper corresponds to the onset of bulk modes. Our study focuses on the gray region in between where only the wall mode is unstable. Note that the studies reported on this plot (symbols) do not all have the same Prandtl number $Pr$ nor the same aspect ratio $\Gamma$.}
\label{fig:param}
\end{center}
\end{figure}

In the present work we present evidence that similarly robust states can be found in forced dissipative hydrodynamics and in particular in {\it nonlinear} systems of this type. While the general theory does not apply to such systems (our system supports multiple boundary currents depending on the parameters used, and these currents may themselves undergo instabilities \citep{zhongES2}), they may nevertheless behave in the same manner as linear systems exhibiting topologically protected boundary currents. Our work is motivated in part by the boundary zonal currents recently observed in high-Rayleigh-number experiments and simulations of rapidly rotating convection in a right circular cylinder rotating with constant angular velocity about its vertical axis \citep{kunnen,bodenschatz}. While the results differ in detail, both groups observe an intense retrograde zonal flow along the cylinder boundary that precesses with respect to the rotating frame and is nearly indifferent to the bulk turbulent state. The emergence of such zonal flows and their coexistence with bulk convection has been known for some time \citep{sanchez2005square,horn2017prograde,aurnou2018rotating}, but the recent experiments are performed at much lower Ekman numbers and larger Rayleigh numbers (see figure~\ref{fig:param}).

In the following we argue that in rotating Rayleigh-B\'enard convection the bulk behaves like a chiral fluid \citep{zhongES} and that, as a result, the presence of a boundary current is not surprising. However, real fluids are dissipative and hence such a boundary current requires the presence of forcing, i.e., a finite Rayleigh number, for its appearance. States of this type were first revealed, indirectly, in experiments by \cite{rossby} and computed, within linear theory, by \cite{goldstein}. The resulting modes, called wall modes to distinguish them from the bulk modes that set in at higher Rayleigh numbers, are confined to the vicinity of the boundary and have been studied in experiments in cylindrical domains \citep{zhongES,zhongEK}. Both modes precess in the rotating frame in a retrograde direction, but the precession rate of the wall modes is in general substantially higher than that of the bulk modes. The wall modes set in supercritically \citep{zhongEK}, and may coexist with other wall modes with different azimuthal wavenumbers \citep{zhongES2}; they may also undergo modulational instabilities of Eckhaus-Benjamin-Feir type as the applied Rayleigh number increases \citep{zhongES2,liu_ecke}. Despite this, we show here that these states bear all the hallmarks of topologically protected states. We first show -- in a cylindrical geometry -- that these states are robust with respect to secondary instabilities and to a turbulent bulk state as in the experiments \citep{kunnen,bodenschatz} by gradually increasing the Rayleigh number until the bulk becomes turbulent. Second, we show that they persist in the presence of different types of barriers, and happily follow whatever boundary geometry is provided. Finally, we show that they persist in the presence of both types of perturbations together, i.e., when barriers are present and the bulk is turbulent.  

\section{Formulation}

We consider the evolution of an incompressible fluid inside a vertical right circular cylindrical container of diameter $D$ and height $H$. The lower horizontal plate is maintained at a higher temperature than the upper plate, while the circular walls are assumed to be thermally insulating. All walls are impenetrable and no-slip. The layer rotates anticlockwise about the $z$ axis, pointing vertically upwards, with a constant angular velocity $\bm{\Omega}=\Omega\bm{e}_z$, while gravity points downwards, $\bm{g}=-g\bm{e}_z$. The kinematic viscosity $\nu$ and thermal diffusivity $\kappa$ are assumed to be constant.

In the Boussinesq approximation, using the rotation time $1/(2\Omega)$ as a unit of time and the depth $H$ of the layer as a unit of length, the dimensionless equations are
\begin{equation}
\label{eq:momentum}
\frac{\partial\bm{u}}{\partial t}+\bm{u}\cdot\nabla\bm{u}=-\nabla p-\bm{e}_z\times\bm{u}+\frac{Ra E^2}{Pr}\theta\bm{e}_z+E\nabla^2\bm{u} \ ,
\end{equation}
\vspace{-6mm}
\begin{equation}
\label{eq:div}
\nabla\cdot\bm{u}=0 \ ,
\end{equation}
\vspace{-6mm}
\begin{equation}
\label{eq:temp}
\frac{\partial\theta}{\partial t}+\bm{u}\cdot\nabla\theta=u_z+\frac{E}{Pr}\nabla^2\theta \ .
\end{equation}
Here $\bm{u}=(u_x,u_y,u_z)$ is the velocity, $p$ is the pressure and $\theta$ is the temperature fluctuation relative to the conduction profile, which varies linearly with respect to the vertical coordinate $z$.
The control parameters are the Rayleigh number $Ra\!=\!\alpha g \Delta TH^3/(\nu\kappa)$, the Ekman number $E\!=\!\nu/(2\Omega H^2)$ and the Prandtl number $Pr\!=\!\nu/\kappa$.
These dimensionless quantities involve $\alpha$, the coefficient of thermal expansion, and $\Delta T$, the imposed temperature difference between the two horizontal plates.
The aspect ratio $\Gamma\!\equiv\!D/H$ specifies the geometry.

We solve equations~\eqref{eq:momentum}-\eqref{eq:temp} using the open-source spectral-element code Nek5000\footnote{NEK5000 Version 19.0. Argonne National Laboratory, Illinois.
Available: \url{https://nek5000.mcs.anl.gov}.} \citep{fischer1997overlapping}. The mesh is composed of up to $19 968$ hexahedral elements and we use a polynomial order up to $N=16$ including dealiasing. The mesh is refined close to the boundaries of the domain in order to properly resolve both Ekman and Stewartson layers. Numerical convergence of the results has been checked by increasing the polynomial order.

\section{Nonlinear dynamics of wall modes\label{sec:main}}

We are interested in the dynamics of the wall modes, which are the first unstable modes in laterally bounded rotating Rayleigh-B\'enard convection. The critical Rayleigh number for the appearance of wall modes can be found in \cite{busse1993} and \cite{kuo1993} (see \cite{zhang_liao_2009} for higher-order corrections) and is given in our dimensionless units by $Ra_c^{\textrm{wall}}\approx\pi^2(6\sqrt{3})^{1/2}E^{-1}$,
while the critical Rayleigh number for the onset of the bulk mode is given by $Ra_c^{\textrm{bulk}}\approx3(\pi^2/2)^{2/3}E^{-4/3}$ \citep{chan61,clune1993,liao_zhang_chang_2006}.
These expressions are valid for low Ekman numbers and are independent of both the Prandtl number and aspect ratio provided the latter is sufficiently large. Although formally obtained for different boundary conditions at the top and bottom, the analysis of \cite{clune1993} explains why the leading-order behavior is independent of these boundary conditions. Both expressions are shown in figure~\ref{fig:param}, where we also report previous experimental and numerical studies of interest. Interestingly, the range of Rayleigh numbers for which only wall modes are unstable increases as the Ekman number decreases.
The above asymptotic results belie the fact that the onset wavenumber is $m=(\pi/2)\Gamma(2+\sqrt{3})^{1/2}\approx 3\Gamma$ as $E\to 0$ \citep{zhang_liao_2009}. Since $m$ is necessarily an integer, each $m$ defines a neutral stability curve, and the intersection between two successive curves heralds a transition from onset wavenumber $m$ to onset wavenumber $m+1$ as $\Gamma$ increases. Thus the critical Rayleigh number is in fact an oscillatory function of $\Gamma$, although its minima are independent of $\Gamma$ for large $\Gamma$.

\begin{figure}
\begin{center}
(a)\includegraphics[height=0.275\textwidth]{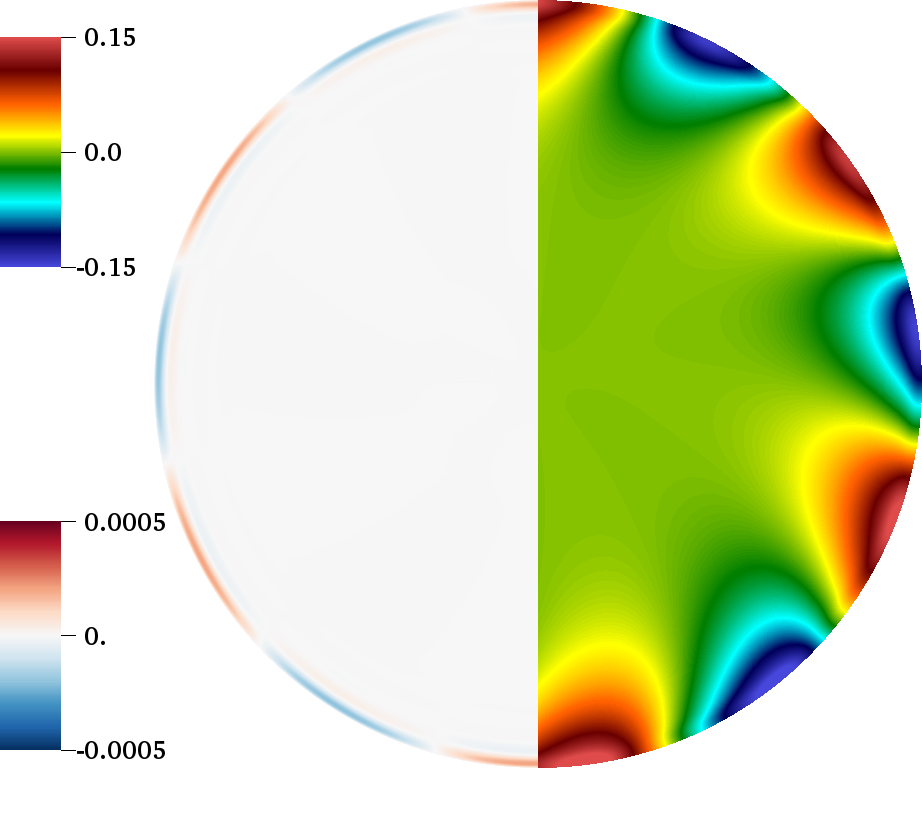}\hfill\includegraphics[height=0.275\textwidth]{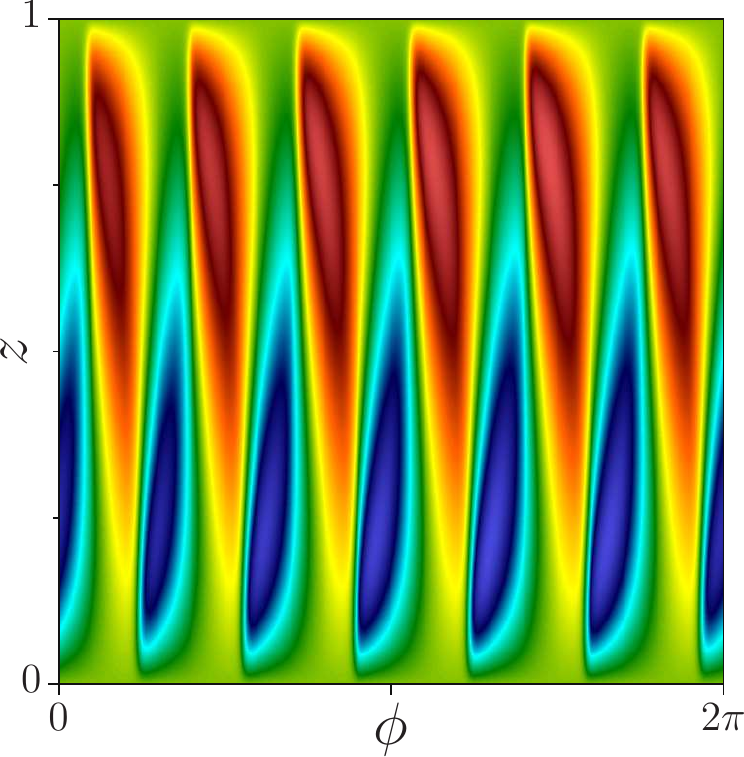}\hfill\includegraphics[height=0.275\textwidth]{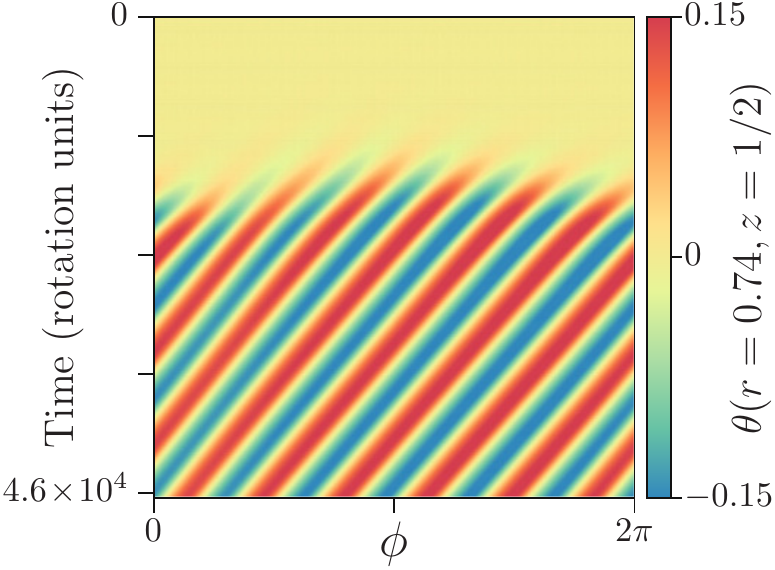}\\

\vspace{3mm}
(b)\includegraphics[height=0.275\textwidth]{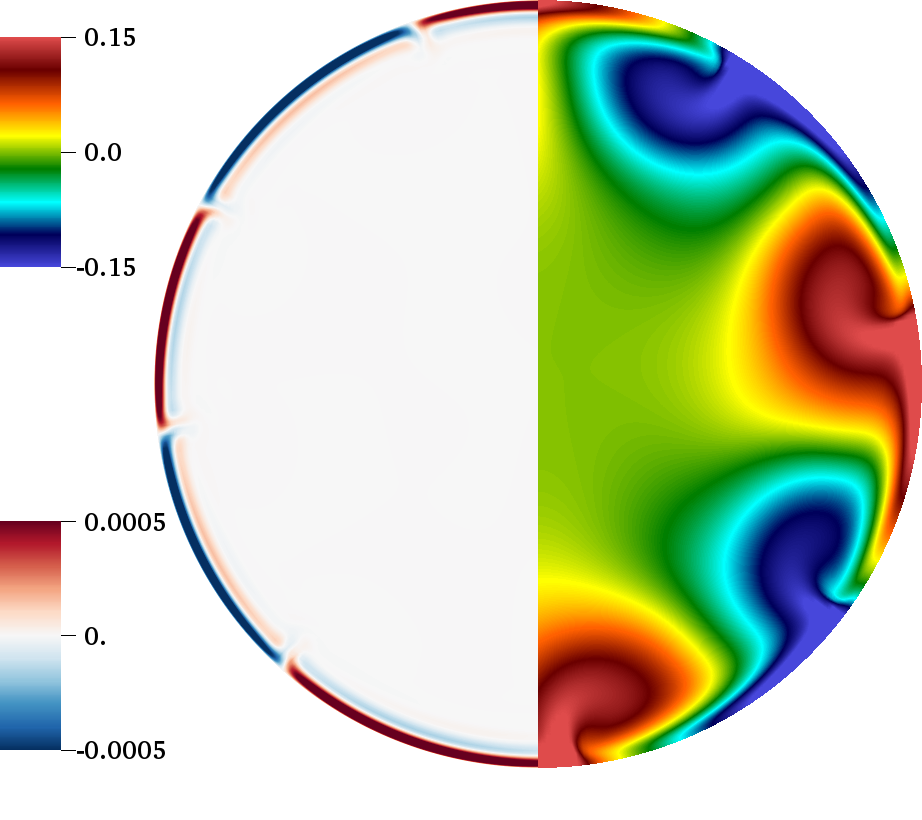}\hfill\includegraphics[height=0.275\textwidth]{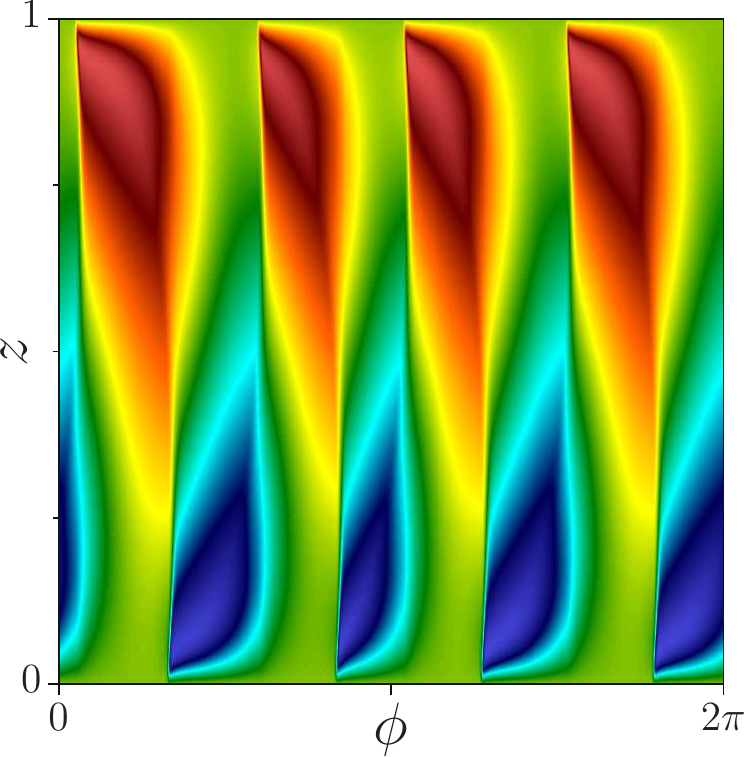}\hfill\includegraphics[height=0.275\textwidth]{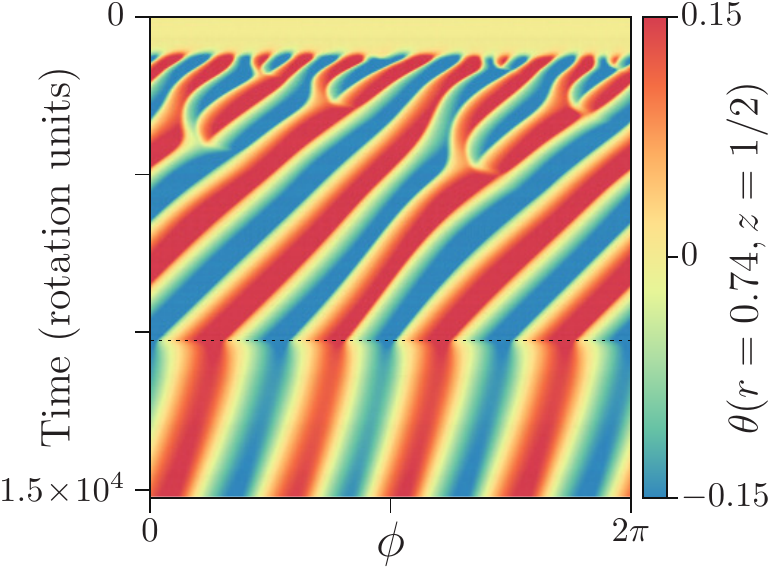}\\

\vspace{3mm}
(c)\includegraphics[height=0.275\textwidth]{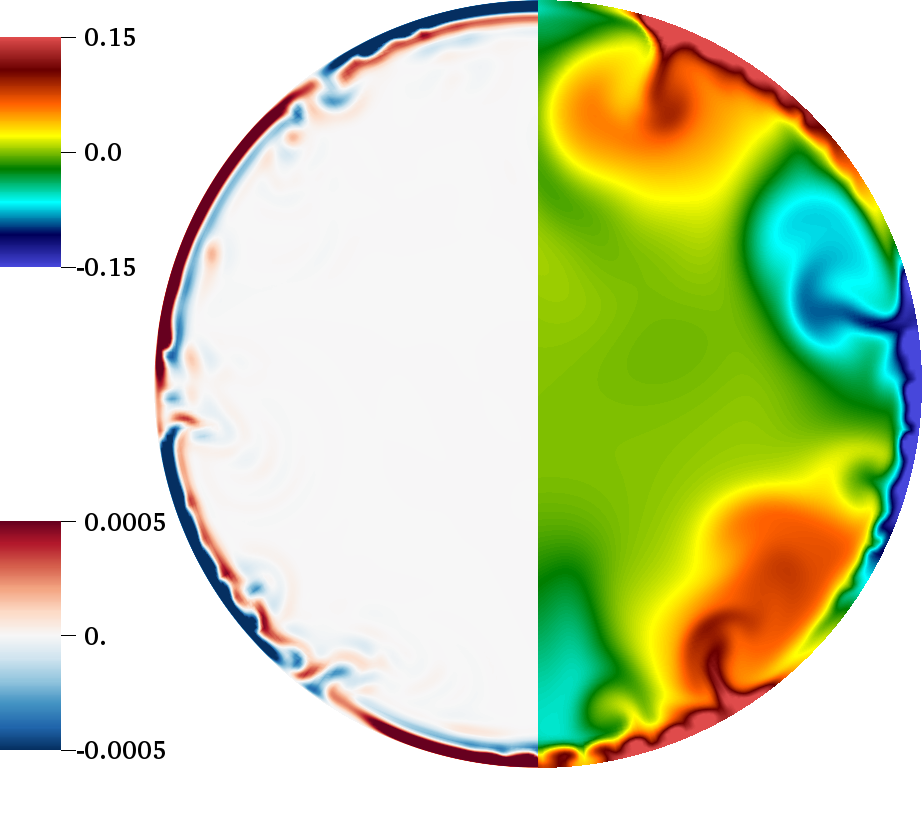}\hfill\includegraphics[height=0.275\textwidth]{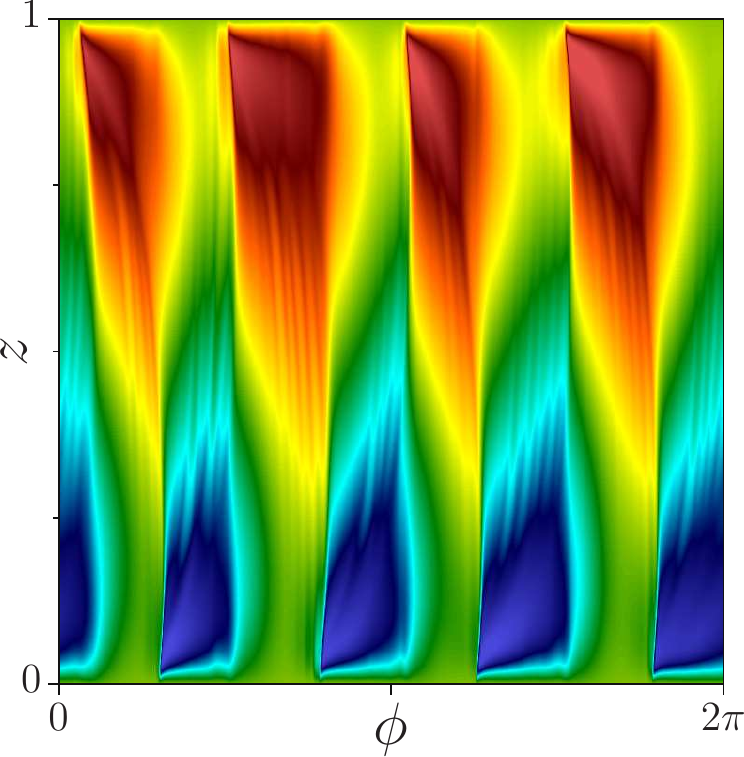}\hfill\includegraphics[height=0.275\textwidth]{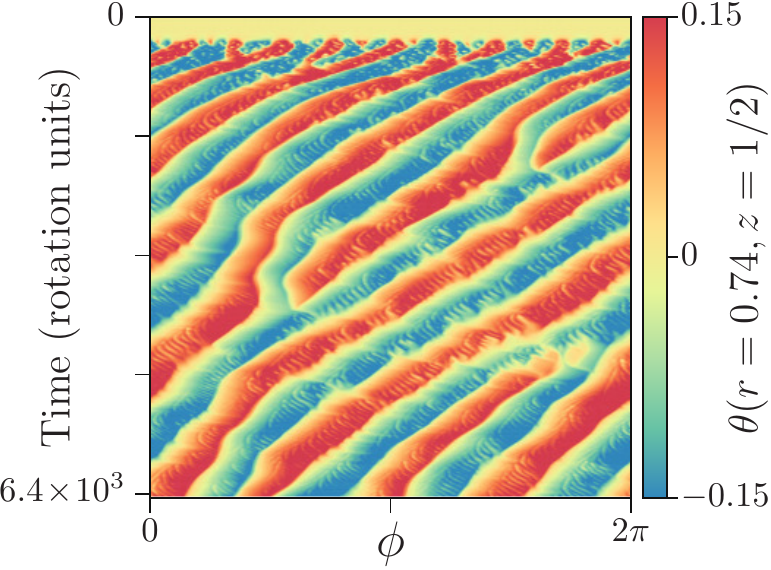}\\

\vspace{3mm}
(d)\includegraphics[height=0.275\textwidth]{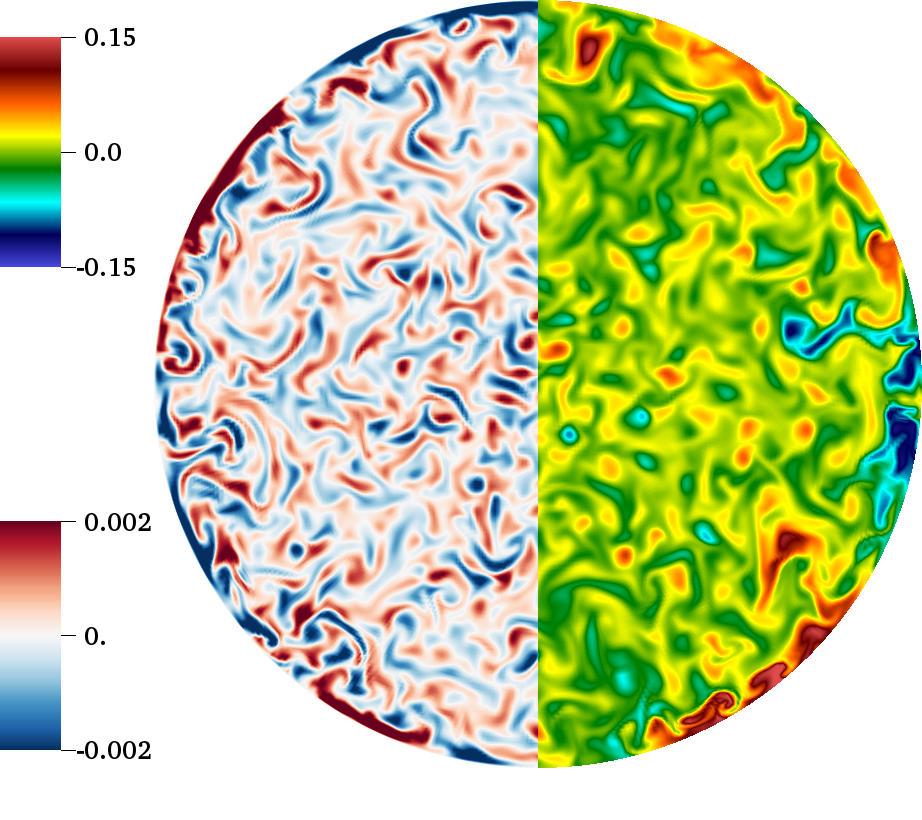}\hfill\includegraphics[height=0.275\textwidth]{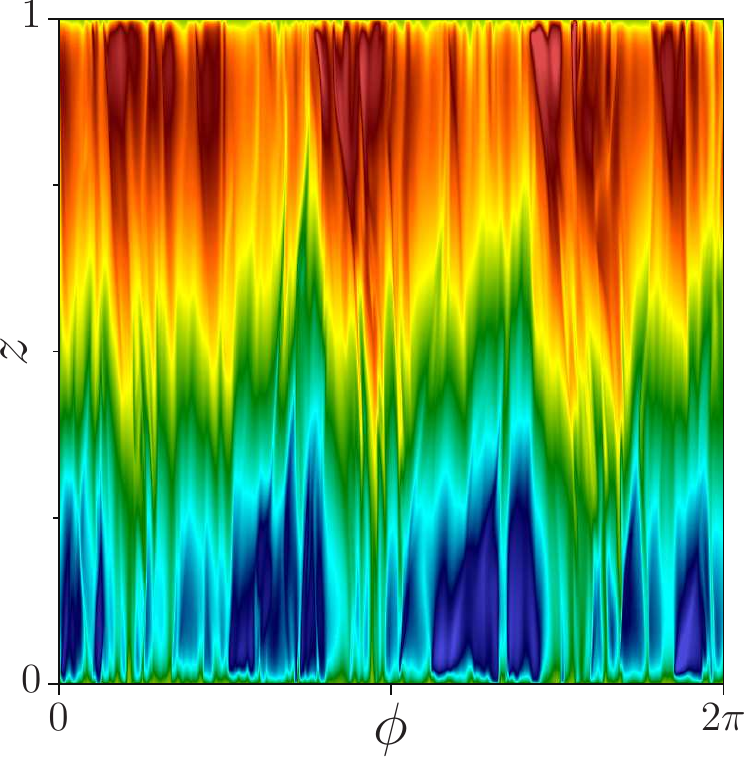}\hfill\includegraphics[height=0.275\textwidth]{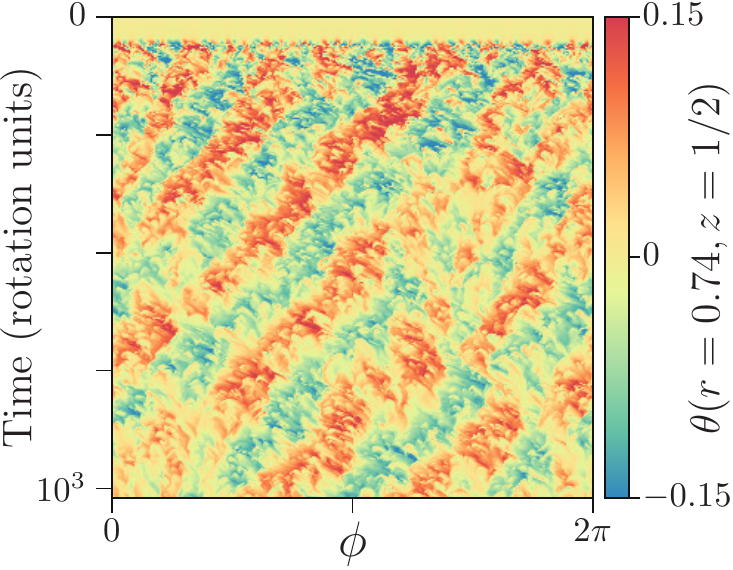}\\
\caption{\underline{Left:} visualizations of the instantaneous vertical velocity $u_z$ (left side) and fluctuating temperature $\theta$ (right side) in the mid-plane $z=0.5$ in a $\Gamma=1.5$ cylinder in the quasi-stationary state. \underline{Middle}: instantaneous temperature fluctuations $\theta$ on the side wall $r=\Gamma/2$ as a function of the azimuthal angle $\phi$ and vertical coordinate $z$ in the quasi-stationary state. \underline{Right:} spatio-temporal plots showing the evolution of the temperature fluctuation $\theta$ at $z=0.5$ and radius $r=0.74$, with $\phi$ plotted horizontally and time (in rotation units) increasing downwards. The time interval is different in each panel. In each case, the simulation is run for at least a hundred free-fall times, which corresponds to thousands of rotation times. The parameters are $\Gamma=1.5$, $E=10^{-6}$ and $Pr=1$; in such a cylinder the wall is at $r=0.75$. The Rayleigh number increases from top to bottom: (a) $Ra=5\times 10^7$, (b) $Ra=2\times 10^8$, (c) $Ra=5\times 10^8$ and (d) $Ra=2\times 10^9$. In panel (b), the dashed horizontal line indicates a reduction of $Ra$ from $2\times10^8$ to $5\times10^7$: the $m=4$ mode survives albeit with a slower precession frequency.}
\label{fig:hovm}
\end{center}
\end{figure}

In this section, we fix $E=10^{-6}$ so that $Ra_c^{\textrm{wall}}\approx3.2\times10^7$ while $Ra_c^{\textrm{bulk}}\approx8.7\times10^8$. We then vary the Rayleigh number from $Ra=5\times10^7$ to $Ra=8\times10^9$ to explore the onset of the wall mode, its nonlinear saturation and finally its coexistence with the bulk mode. The aspect ratio is $\Gamma=1.5$ and the Prandtl number is $Pr=1$ for simplicity (see section~\ref{sec:exp} for parameters relevant to the recent experiments).
The simulations are initialized with small-amplitude temperature fluctuations. In each case we observe exponential growth followed by nonlinear saturation. Visualizations of the vertical velocity and the temperature fluctuation in the mid-plane $z=0.5$ in the saturated state are shown in the left panels of figure~\ref{fig:hovm}. At $Ra=5\times 10^7$, i.e. relatively close to the onset of the wall mode, we find a retrograde $m=6$ travelling wave localized near the sidewall (figure~\ref{fig:hovm}(a)). The travelling nature of the wall mode is reflected in the asymmetry of its instantaneous profile (see the temperature fluctuations on the side boundary in the middle panels of figure~\ref{fig:hovm}), but is best seen by looking at spatio-temporal diagrams of the temperature fluctuation in the mid-plane $z=0.5$ and an arbitrary location $r$ close to the boundary: $r=0.74<\Gamma/2=0.75$ (right panels in figure~\ref{fig:hovm}). The results at other radii and depths are very similar provided one remains sufficiently close to the outer boundary. This travelling mode is similar to that observed in experiments in a $\Gamma=2$ cylinder rotating at lower rotation rates \citep{zhongES,zhongES2}. The structure of this mode is also similar to that of the critical mode for $\Gamma=4$, $E=5\times 10^{-5}$ and $Pr=1$ computed by \cite{zhang_liao_2009} for which $m=10$ while the asymptotic wavenumber as $E\to 0$ is $m\approx 12$. In contrast, in our case, computed for $E=10^{-6}$, the observed wavenumber is larger than the onset theoretical wavenumber $m\approx 4.5$. Evidently our choice of parameters is outside the asymptotic regime $E\to0$ and $\Gamma\to\infty$.

As the Rayleigh number increases -- while remaining below the threshold for the bulk mode -- the wall mode becomes fully nonlinear and more complex dynamical behaviour is seen. We first observe (see figure~\ref{fig:hovm}(b)) an Eckhaus-type instability \citep{janiaud1992,knobloch1994}, as found experimentally at much higher Ekman number by \cite{liu_ecke}. This instability leads to the gradual coarsening or merging of adjacent cells but the flow remains laminar. The transitions between different azimuthal wavenumbers are expected to be strongly hysteretic for reasons explained in \cite{knobloch1994}, a behaviour that is illustrated in figure~\ref{fig:hovm}(b), where the $m=4$ wall mode persists even when the Rayleigh number is reduced from $2\times 10^8$ back to $5\times 10^7$, i.e., to a Rayleigh number at which $m=6$ is the first mode that appears.
At even higher Rayleigh numbers ($Ra=5\times10^8$, see figure~\ref{fig:hovm}(c)), the $m=4$ travelling wave pattern persists but becomes fully nonlinear and starts to emit plumes into the stable bulk resulting in short-wavelength perturbations of the mode. These fluctuations tend to propagate in the opposite direction from that of the travelling wave, i.e., in a cyclonic sense. Finally, a $m=3$ wall mode survives even for Rayleigh numbers beyond the onset of bulk convection, where it takes the form of a nonlinear wall-confined travelling wave superposed on small-scale bulk turbulence (figure~\ref{fig:hovm}(d)).

\begin{figure}
\begin{center}
\includegraphics[height=0.4\textwidth]{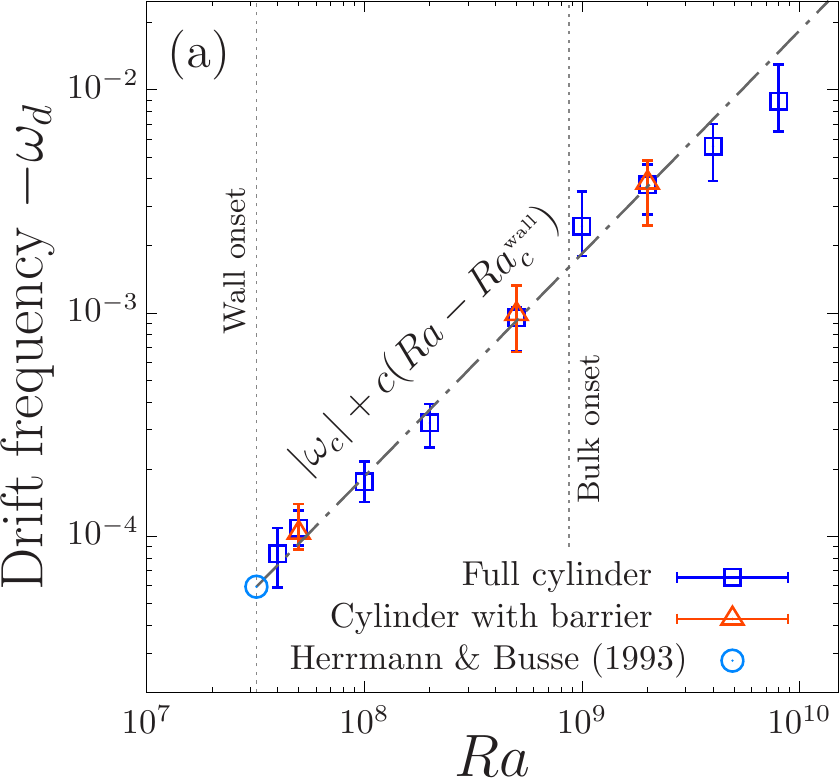}\hspace{7.2mm}\includegraphics[height=0.4\textwidth]{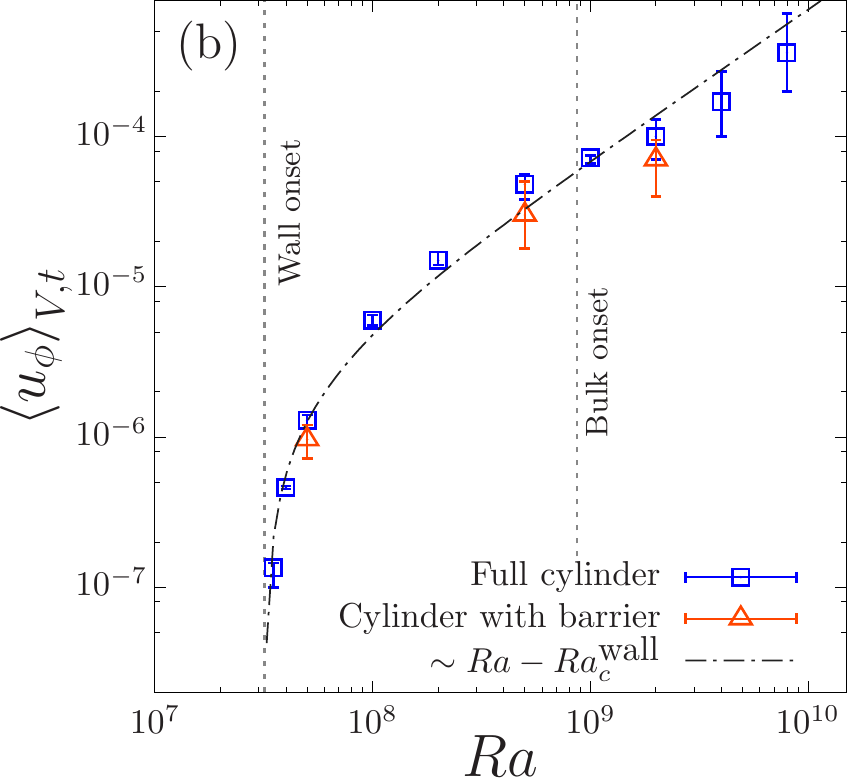}\\
\vspace{4mm}
\hspace{2mm}\includegraphics[height=0.4\textwidth]{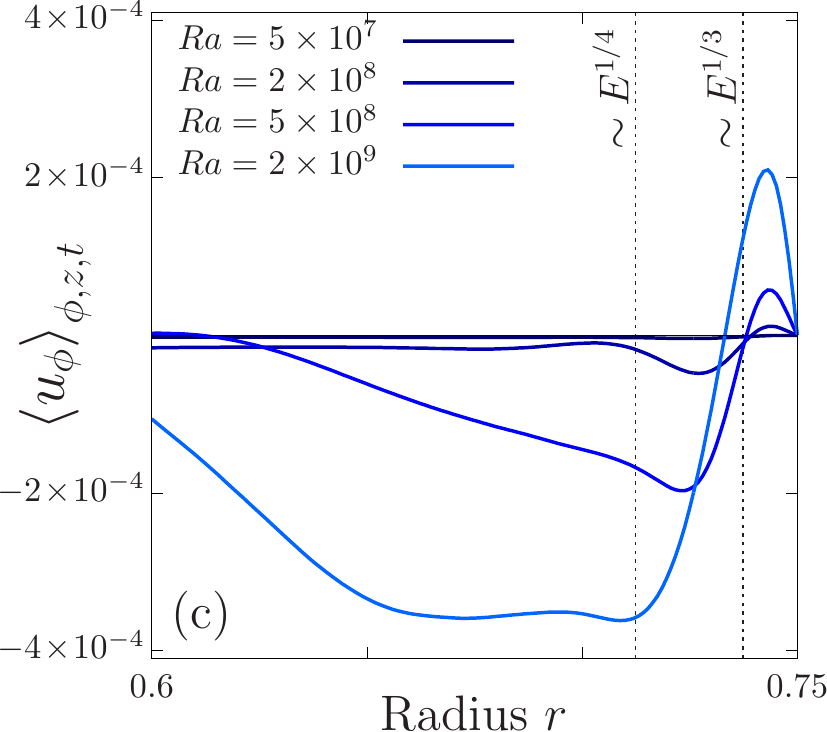}\hspace{6mm}\includegraphics[height=0.4\textwidth]{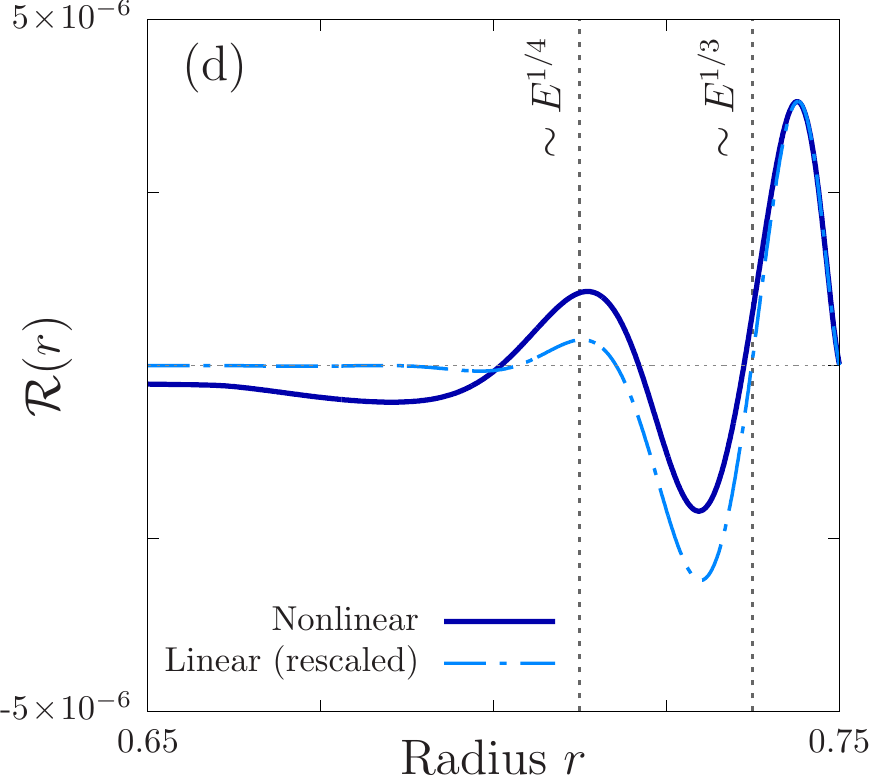}
  \caption{(a) Drift frequency $-\omega_d$ as a function of $Ra$ for $\Gamma=1.5$, $E=10^{-6}$ and $Pr=1$. The results for the full cylinder (\textcolor{blue}{$\square$}) and the cylinder with a barrier (\textcolor{orange}{$\triangle$}, see section~\ref{sec:barrier} below) coincide. The theoretical value $\omega_c\approx-59 E/Pr$ predicted by \cite{busse1993} for the onset of the instability in the presence of a planar wall is also reported (open circle). The dot-dash line corresponds to the linear scaling $|\omega_d|=|\omega_c|+c(Ra-Ra_c^{\textrm{wall}})$ with $c$ an arbitrary constant. (b) Corresponding volume- and time-averaged zonal velocity $\left<u_{\phi}\right>_{V,t}$. The dot-dash line corresponds to the scaling $\left<u_{\phi}\right>_{V,t}\sim Ra-Ra_c^{\textrm{wall}}$. The volume integration is only performed over the left half of the cylinder for the cases with barrier (see section~\ref{sec:barrier}). (c) Azimuthally, vertically and temporally averaged zonal velocity $\left<u_{\phi}\right>_{\phi,z,t}$ as a function of the radial coordinate $r$. Positive values correspond to cyclonic motions while negative values correspond to anticyclonic motions. The two vertical lines indicate the Stewartson layer scales $E^{1/3}$ and $E^{1/4}$ \citep{stewartson_1957}. (d) Divergence of the Reynolds stress in the azimuthal direction for the case $Ra=5\times10^8$. Results are computed at a given time during the exponential phase of the instability (arbitrarily rescaled) and time-averaged during the nonlinear saturated phase.}
\label{fig:drift}
\end{center}
\end{figure}

From the spatio-temporal diagrams one can easily extract the drift frequency $\omega_d$ as a function of the Rayleigh number. To do so, we perform a temporal Fourier transform of the fluctuating temperature at a given point and extract the frequency corresponding to the maximum of the power spectrum. This frequency is shown in figure~\ref{fig:drift}(a) (all values are negative so we plot $-\omega_d$). The drift frequency increases continuously with increasing $Ra$ and connects to the theoretical prediction of \cite{busse1993} at onset, $\omega_c\!\approx\!-59E/Pr$. This increase appears to follow a linear scaling with the Rayleigh number, $|\omega_d-\omega_c|\!\sim\!Ra-Ra_c^{\textrm{wall}}$, as observed at higher Ekman numbers by \cite{horn2017prograde} and more recently by \cite{kunnen} (see section~\ref{sec:exp} below). In fact such a linear relationship between the wall mode drift frequency and the Rayleigh number is expected close to the Hopf bifurcation from which the mode originates \citep{zhongEK,goldstein}, but it is unexpected that it continues to hold over nearly two decades in Rayleigh number, and extends into the regime where the bulk mode is unstable and even turbulent. This is a first indication of the robustness of the wall state.

All wall states can be decomposed into an axisymmetric component with wavenumber $m=0$ and a drifting non-axisymmetric component with wavenumbers $m>0$. We identify the former with the observed zonal flow \citep{sanchez2005square,kunnen,bodenschatz}, and use the latter to identify the drift frequency $\omega_d$. To study the zonal flow, we compute the time and volume average of the azimuthal velocity component $u_{\phi}$ in the quasi-stationary state (restricting the volume average to a region near the sidewall leads to qualitatively similar results). As shown in figure~\ref{fig:drift}(b), the mean zonal flow amplitude also increases linearly with $Ra-Ra_c^{\textrm{wall}}$ and becomes vanishingly small close to onset. This observation is consistent with the fact that the Nusselt number also increases linearly with the critical parameter near onset (not shown; see \cite{zhongES2} and \cite{horn2017prograde}). Note that the linear scaling once again extends over the whole range of Rayleigh numbers considered here, including cases with and without bulk turbulence, with no significant impact of the transition between them. This scaling is, of course, not expected to hold indefinitely as the Rayleigh number is increased even further; the flow becomes less rotationally constrained and contributions from the bulk eventually dominate.

Following \cite{kunnen}, we determine the radial structure of the zonal flow by computing the azimuthally, vertically and temporally averaged zonal velocity $\left<u_{\phi}\right>_{\phi,z,t}$ as shown in figure~\ref{fig:drift}(c). Close to onset, the mean azimuthal flow is negligible, but with increasing $Ra$ it strengthens rapidly and develops a concentric ring structure, with cyclonic flow in the outer ring and anticyclonic flow in the inner ring. This flow structure is typical of Stewartson layers \citep{stewartson_1957,kunnen_2011}, with the cyclonic mean flow confined to a layer of thickness $E^{1/3}$ close to the boundary and the anticyclonic flow present in a broader layer of thickness $E^{1/4}$, exactly as observed by \cite{bodenschatz} (see their figure~3(b)). Interestingly, one can now compute the typical Reynolds number associated with this mean differential rotation. Based on the largest layer thickness $E^{1/4}$ and the azimuthal mean velocity difference in the profile shown in figure~\ref{fig:drift}(b), one obtains $Re\approx2.3$ at $Ra=2\times10^8$ and $Re\approx9.7$ at $Ra=5\times10^8$. Since the critical Reynolds number for shear barotropic instabilities is typically around $Re\approx10-20$ independently of $E$ \citep{hide_titman_1967,niino,fruh_read_1999}, we conjecture that the small-scale fluctuations observed at $Ra=5\times10^8$ but not at $Ra=2\times10^8$ are driven by a shear instability of the mean barotropic zonal flow \citep{busse_1968}, although a complete analysis of this instability is beyond the scope of this paper. The same nested cyclonic-anticyclonic structure is observed even after the appearance of the bulk state, indicating that this structure is a consequence of the survival of the nonlinear wall state into this regime. The observed spreading of the anticyclonic flow into the bulk as $Ra$ increases, well beyond the linear Stewartson layer, indicates that additional effects come into play. This spreading is observed as soon as the wall mode becomes unstable to secondary instabilities (i.e., for $Ra\gtrsim 5\times10^8$), even before the bulk mode sets in, and is therefore a property of the nonlinear wall state that cannot be fully attributed to the emergence of the bulk state.

\begin{figure}
\begin{center}
(a)\includegraphics[height=0.29\textwidth]{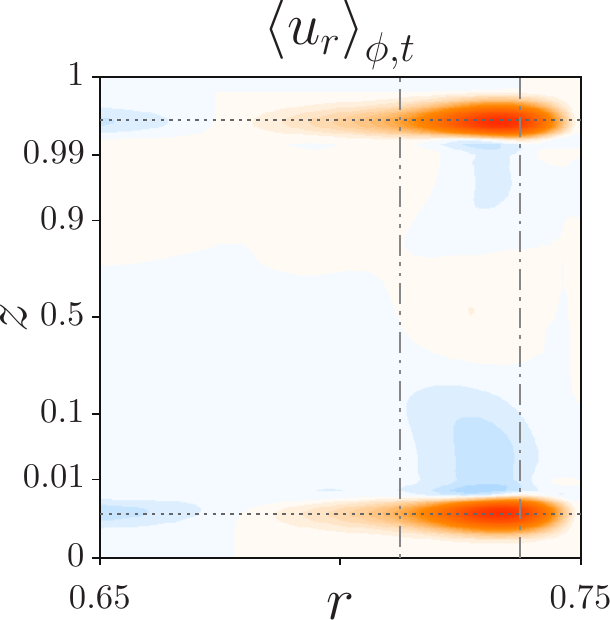}\hfill\includegraphics[height=0.29\textwidth]{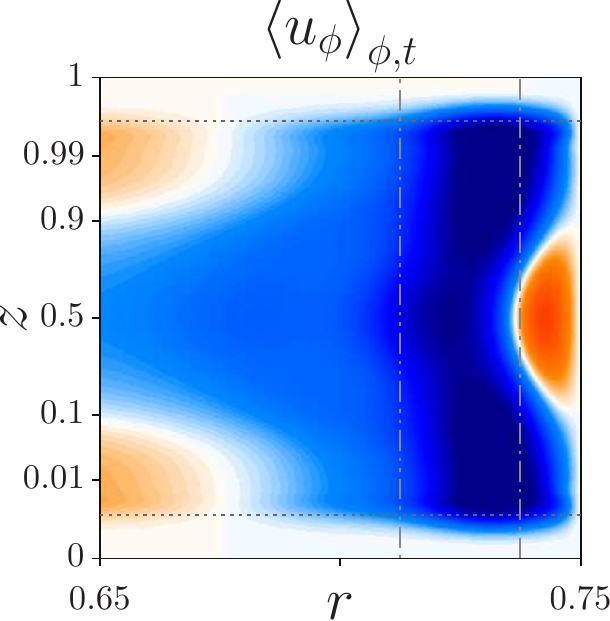}\hfill\includegraphics[height=0.29\textwidth]{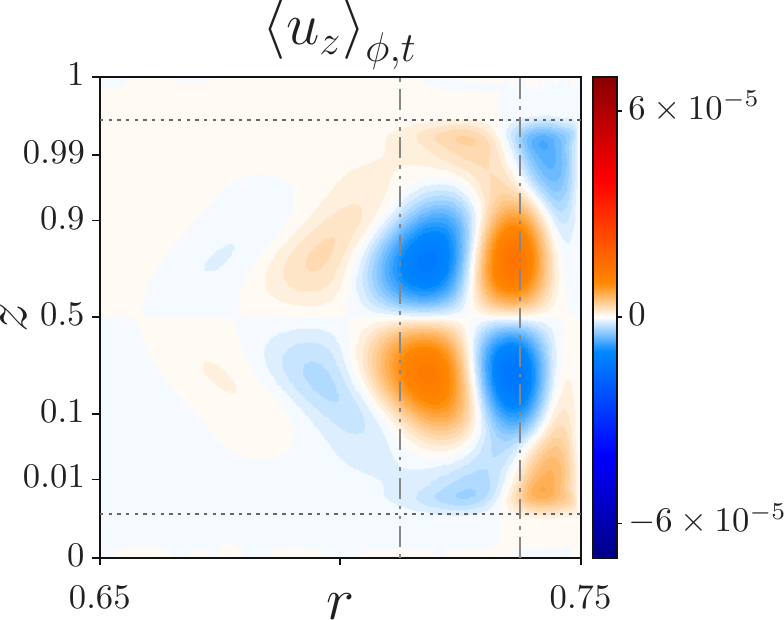}\\
(b)\includegraphics[height=0.26\textwidth]{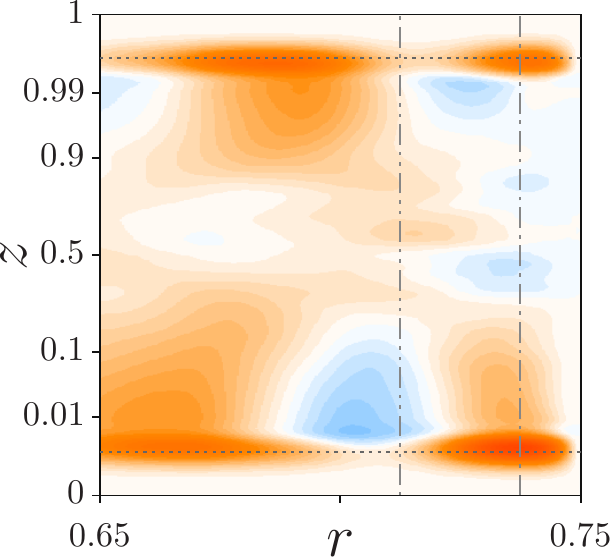}\hfill\includegraphics[height=0.26\textwidth]{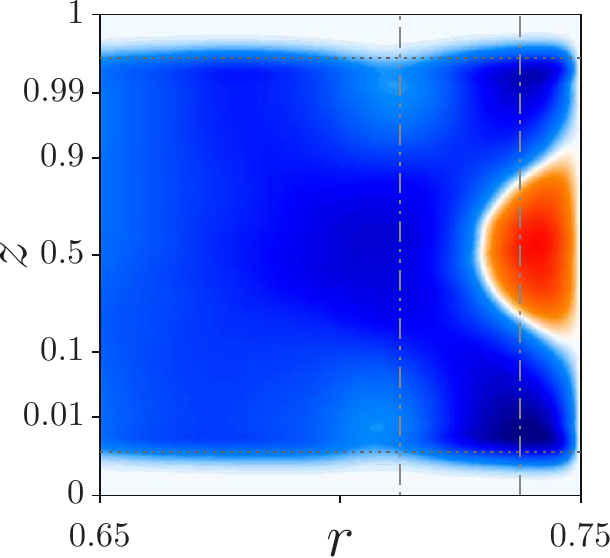}\hfill\includegraphics[height=0.26\textwidth]{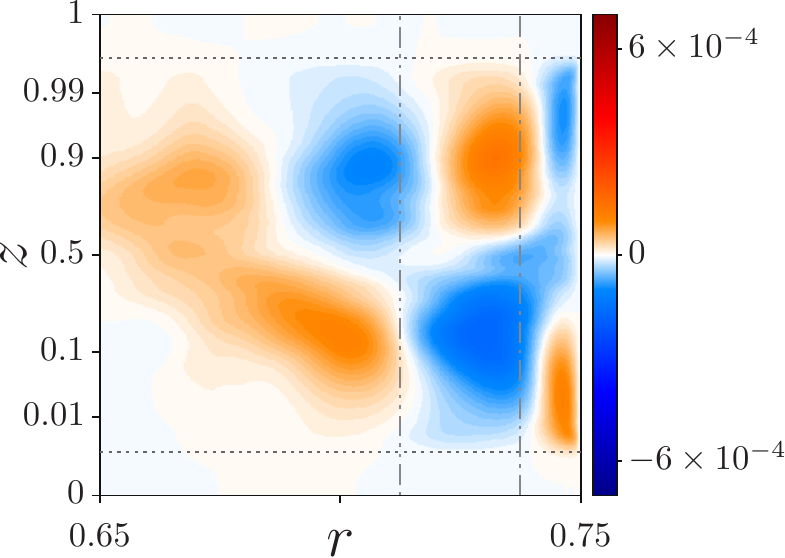}
  \caption{Temporal and azimuthal averages of all three velocity components in cylindrical coordinates as a function of the radial and vertical positions. Plots focus on the region close to the sidewall (located at $r=\Gamma/2=0.75$) and the vertical coordinate is stretched to zoom into the upper and lower walls (the mapping $z\rightarrow\sin^2(\pi z/2)$ is applied twice). The horizontal dotted lines show the vertical thickness $E^{1/2}$ of the Ekman boundary layers while the vertical dashed lines correspond to the Stewartson layer scaling $E^{1/4}$ and $E^{1/3}$. (a) $Ra=2\times10^8$. (b) $Ra=2\times10^9$. Color scales are chosen consistently with the linear scaling observed in figure~\ref{fig:drift}(b).}
\label{fig:mean_flow}
\end{center}
\end{figure}

The mean zonal flow is driven by the non-axisymmetric structure of the wall state itself. While the drift frequency $\omega_d$ is a consequence of a fundamentally linear mechanism, the zonal flow is driven nonlinearly by Reynolds stresses. These stresses are known to sustain both zonal geostrophic and meridional non-geostrophic flows within a weakly nonlinear analysis \citep{liao2005convection}. We confirm here that this mechanism continues to operate at larger Rayleigh numbers by computing the divergence of the Reynolds stresses in the azimuthally and vertically averaged zonal equation, $\mathcal{R}(r)=-\bm{e}_{\phi}\cdot(\nabla\cdot\left<\bm{u}'\bm{u}'\right>_{\phi,z})$, where $\bm{u}'=\bm{u}-\left<\bm{u}\right>_{\phi,z}$. The results are shown in figure~\ref{fig:drift}(d) for $Ra=5\times10^8$. We see that the Reynolds stress associated with the linear wall mode (\textit{i.e.} calculated during the exponential phase of the instability) is indeed responsible for the cyclonic mean flow within the $E^{1/3}$ Stewartson layer and for the dominant outer anticyclonic zonal flow. The structure of this stress persists into the saturated phase, even though the wall mode is now fully nonlinear. In this quasi-steady state the Reynolds stress is in balance with viscous dissipation. It remains to be seen how the stresses generated by the wall state interact with the bulk state, potentially explaining the expansion of the anticyclonic mean flow observed in figure~\ref{fig:drift}(c) and in \cite{bodenschatz}.

We discuss, finally, the full spatial structure of the mean flow, which is not fully captured by the vertically averaged quantity $\left<u_{\phi}\right>_{\phi,z}$. The azimuthal and temporal average of all three components of the velocity are shown in figure~\ref{fig:mean_flow}. The intricate mean flow structure observed for $Ra=2\times10^8$ persists even after the instability of the bulk mode, at $Ra=2\times10^9$. Although the signal is now perturbed by the small-scale fluctuations driven by the bulk state, the same flow structure is easily recognizable, showing that the mean flow is predominantly driven by the wall mode itself even when bulk turbulence is present. As in figure 3(c), the mean flow pattern spreads out radially with increasing $Ra$, in agreement with \cite{bodenschatz}, and all components of the flow follow the linear scaling with the distance from threshold identified in figure 3(b). Moreover, complex meridional structures are present even when the zonal component dominates the mean flow. For example, the cyclonic motion observed in figure~\ref{fig:drift}(c) is centred near the mid-plane and also spreads inwards as $Ra$ increases. We also recover the meridional circulation predicted by weakly nonlinear theory \citep{liao2005convection} with two counter-rotating cells on either side of the mid-plane. Finally, we observe strong outward radial flow inside the top and bottom Ekman layers, a direct consequence of Ekman pumping due to the differential rotation above and below.

\section{Link with recent experiments and simulations at high $Ra$\label{sec:exp}}

\begin{figure}
\begin{center}
(a)\includegraphics[height=0.27\textwidth]{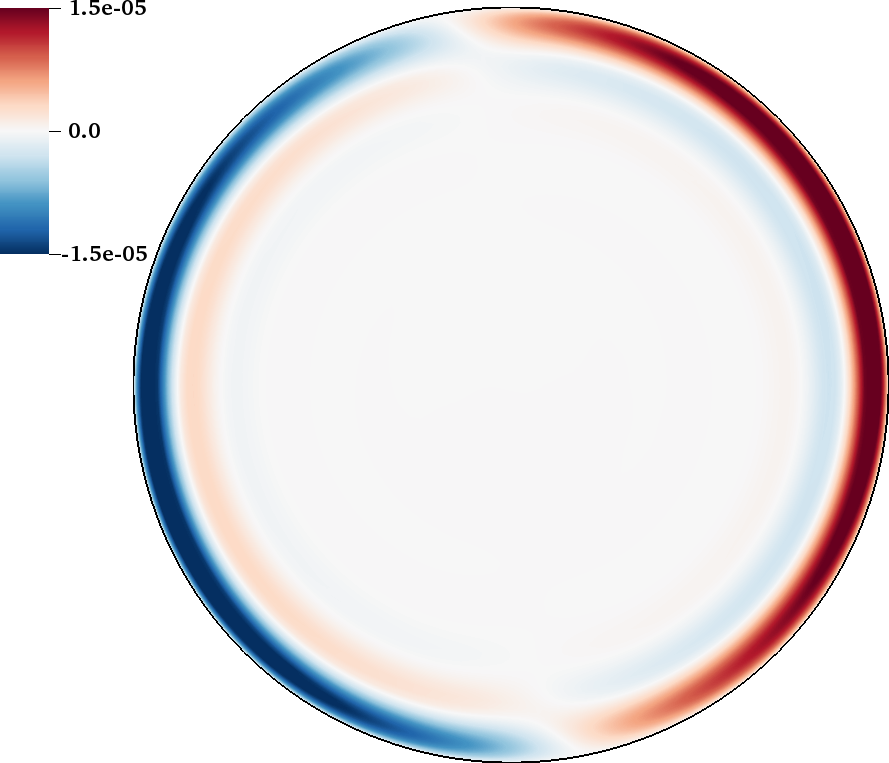}(b)\includegraphics[height=0.27\textwidth]{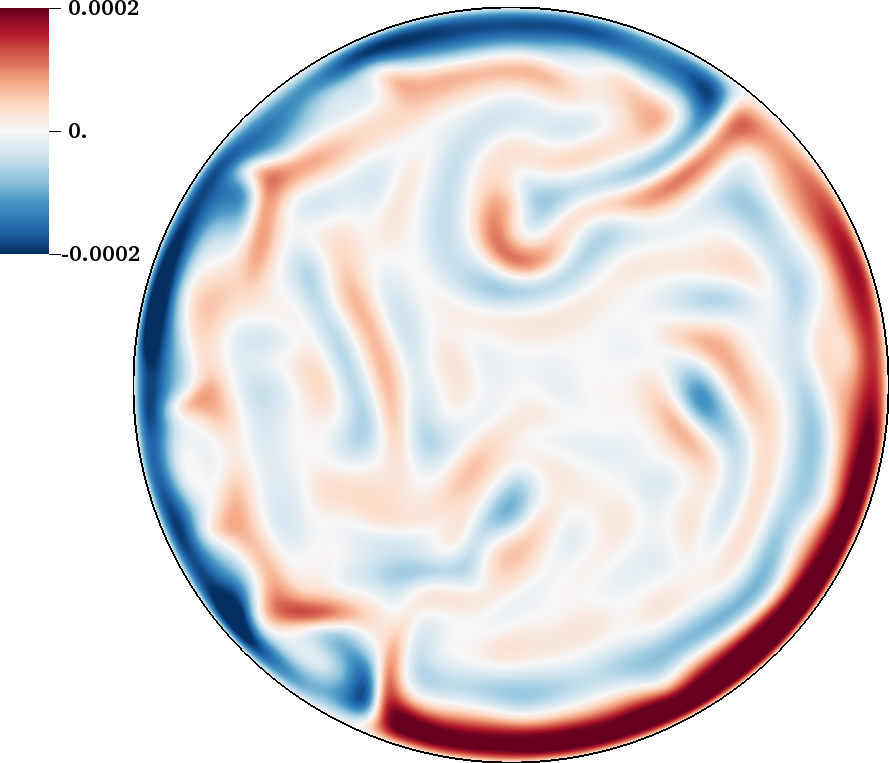}(c)\includegraphics[height=0.27\textwidth]{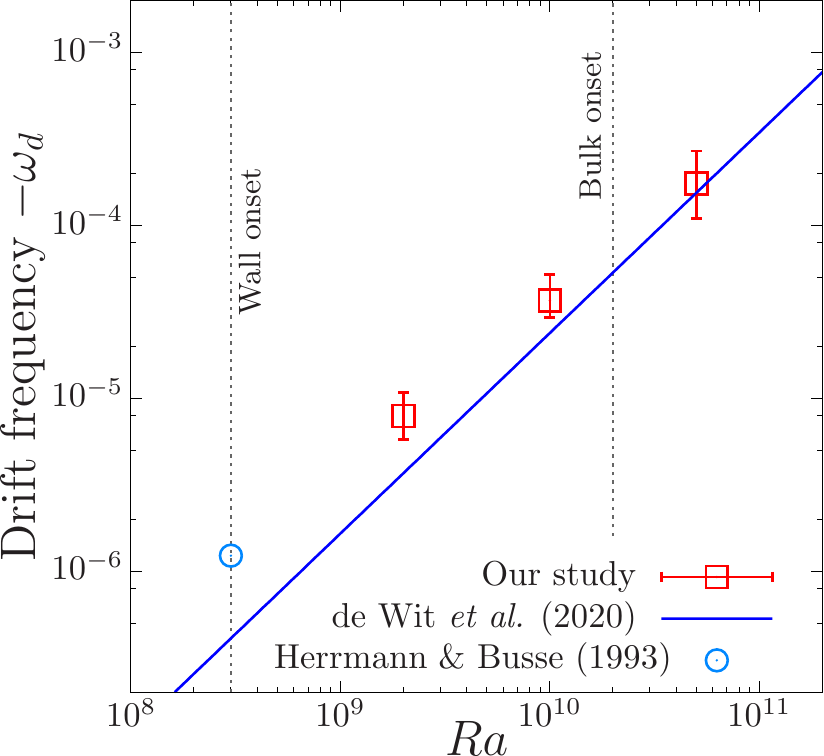}
\caption{Vertical velocity in the mid-plane $z=0.5$ in a $\Gamma=0.2$ cylinder rotating at $E=10^{-7}$ with $Pr=5.2$ and (a) $Ra=2\times10^9$, (b) $Ra=5\times10^{10}$ (this last case can be compared with figure 2(b) of \cite{kunnen} at the exact same parameters). (c) Drift frequency $-\omega_d$ as a function of $Ra$. The theoretical value $\omega_d=\omega_c\approx-63 E/Pr$ predicted by \cite{busse1993} for the onset of the instability in the presence of a planar wall is also reported (open circle). The oblique line corresponds to the scaling $\omega_d\approx-6\times10^{-10}Ra^{1.16}E$ reported in \cite{kunnen} and obtained at much larger $Ra$.}
\label{fig:exp}
\end{center}
\end{figure}

Recent experiments and simulations have shown the emergence of a boundary zonal flow in confined rotating Rayleigh-B\'enard convection \citep{kunnen_2011,kunnen,bodenschatz}. However, a link between this flow and the wall modes was not explicitly made. In further support of this link, we repeated the simulation of \cite{kunnen} using the parameters $E=10^{-7}$, $Pr=5.2$ and $\Gamma=0.2$. Instead of focusing on the dynamics of the bulk mode, as in \cite{kunnen}, we consider Rayleigh numbers below the onset of the bulk mode (i.e., for $Ra<Ra_c^{\textrm{bulk}}\approx2\times10^{10}$), but above the onset of the wall mode ($Ra>Ra_c^{\textrm{wall}}\approx3\times10^8$). We show in figure~\ref{fig:exp}(a) the vertical velocity in the mid-plane at $Ra=2\times10^9$, well beyond the onset of wall modes, once the system has reached a statistically steady state. We observe an $m=1$ travelling wave, presumably a consequence of the small value of $\Gamma$ (the asymptotic onset wavenumber for $\Gamma=0.2$ is $m\approx0.6$). This state is very similar to the boundary zonal flow observed by \cite{kunnen} (see their figure 2(b), for example) at the much higher Rayleigh number of $Ra=5\times10^{10}$. Here the nonlinear dynamics of the wall state are clearly constrained by the small aspect ratio used in these simulations compared to the cases discussed in the previous section. However, once the Rayleigh number is increased to $Ra=5\times10^{10}$, we recover the emergence of the bulk mode superposed on the nonlinear wall mode (see figure~\ref{fig:exp}(b), which can be directly compared with figure 2(b) of \cite{kunnen}). Note that \cite{kunnen} also measured the drift frequency as a function of $Ra$ and found the following fit: $\omega_d\approx-6\times10^{-10}Ra^{1.16\pm0.06}E$ (using our rotation units).

Using the same approach as in section~\ref{sec:main}, we measured the drift frequency in our simulations below the onset of the bulk mode, and in figure~\ref{fig:exp}(c) we compare the results with the fit of \cite{kunnen}. The results clearly bridge the gap between the onset theoretical value derived by \cite{busse1993} and the highly turbulent scaling found by \cite{kunnen}, thereby providing additional evidence for the robustness of the linear scaling discussed in section~\ref{sec:main}. Note also that we reproduce many of the results presented in these studies (such as the bimodal temperature distribution and the boundary zonal flow) by considering cases below the onset of the bulk mode. This provides a clear indication that the boundary dynamics observed in these studies are related to the nonlinear saturation of the wall mode and so are not directly driven by the dynamics in the bulk (which is stable in two of our cases). The emergence of the boundary zonal flow is thus best understood as the nonlinear development of linearly unstable wall modes that eventually couple to the bulk dynamics in the manner described in these studies \citep{kunnen_2011,kunnen,bodenschatz}.

\section{Robustness of the wall modes to geometry changes\label{sec:barrier}}

The time-dependent wall modes we have described thus far are linked to the presence of a sidewall. These states behave like topologically protected states and, as shown above, survive even when the Rayleigh number is very far from its critical value, i.e., in the strongly nonlinear regime. The robustness of these states can also be demonstrated by perturbing the geometry of the domain. This has already been done in other similar systems \citep{dasbiswas,souslov} and was in fact suggested in the very early studies of wall modes in rotating convection by \cite{liu_ecke}; see \cite{vasil2008new} for a related study of drifting states in a domain of square cross-section.

\begin{figure}
\begin{center}
(a)\includegraphics[width=0.31\textwidth]{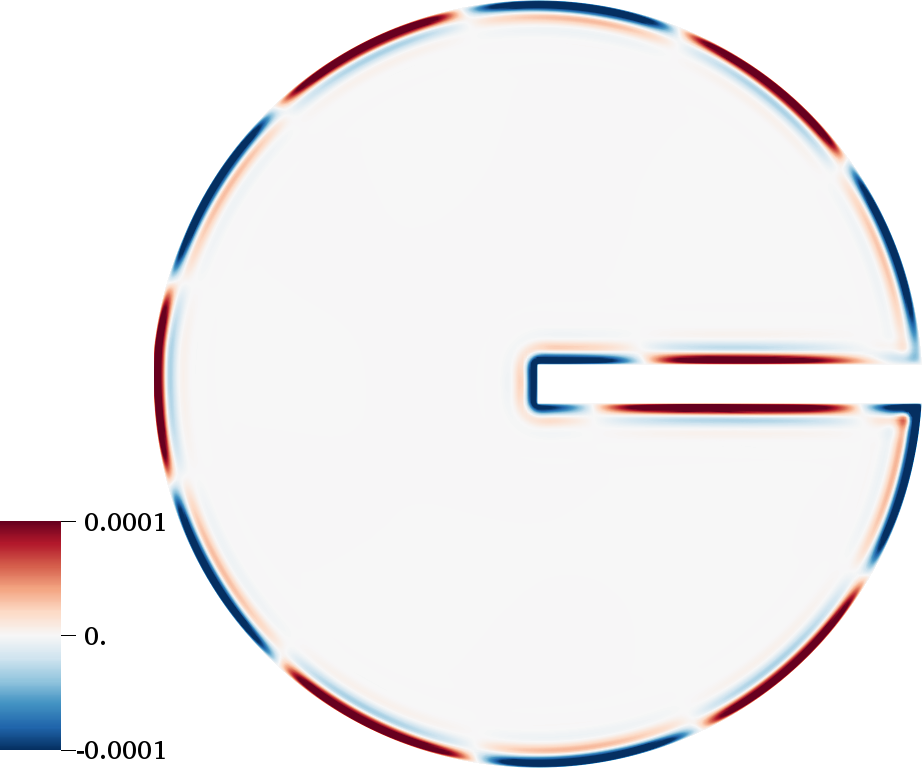}(b)\includegraphics[width=0.31\textwidth]{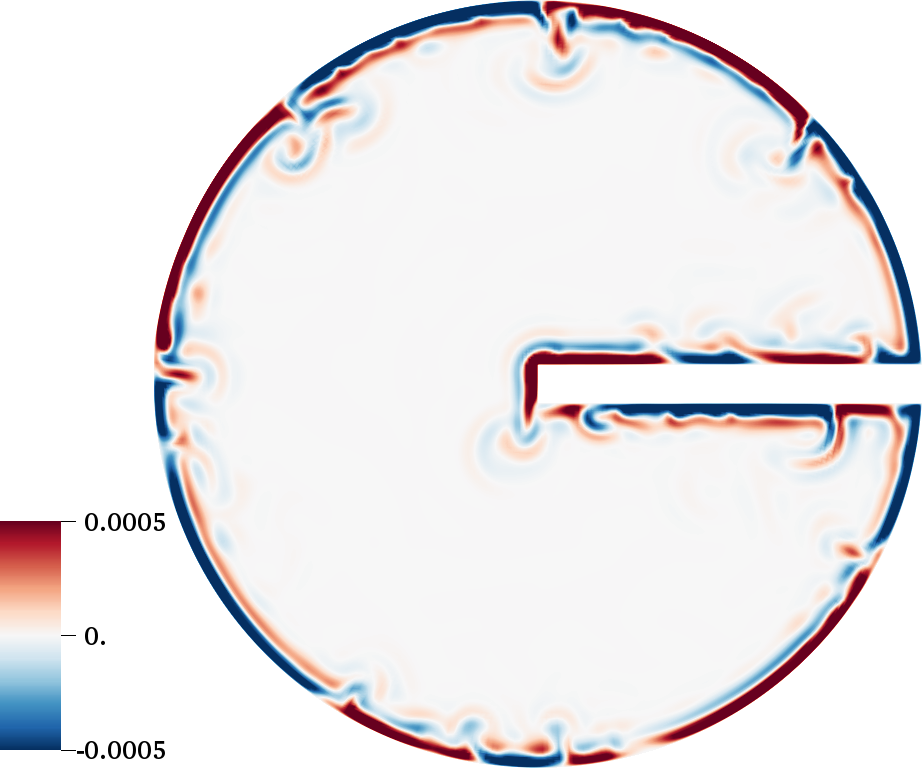}(c)\includegraphics[width=0.31\textwidth]{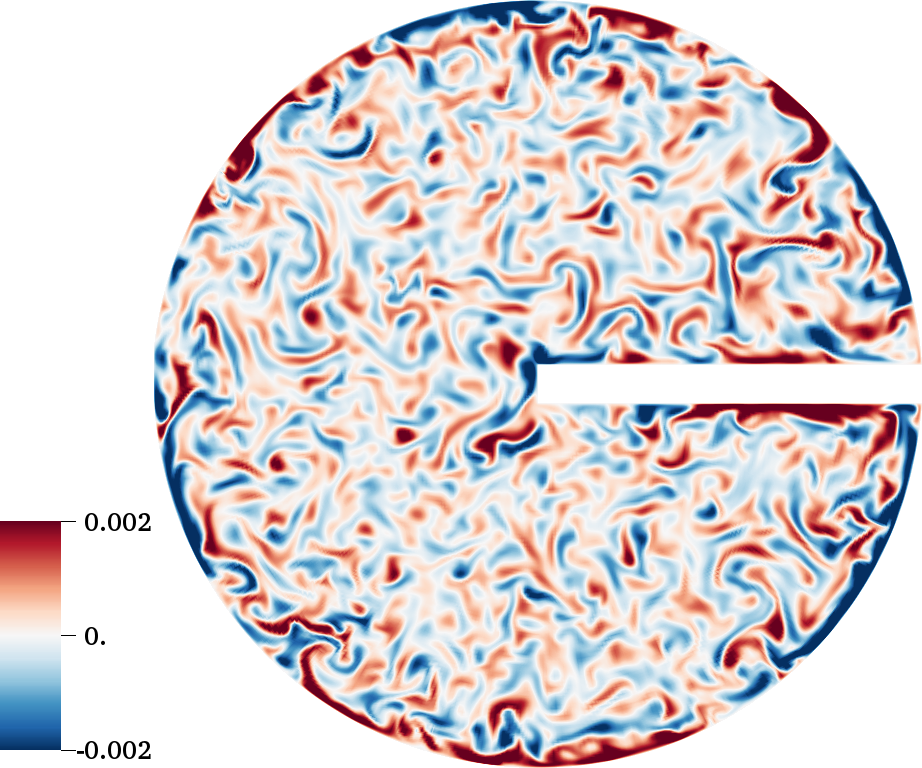}
\caption{Vertical velocity in the mid-plane $z=0.5$ for a cylinder with a barrier. The Rayleigh number increases from left to right: (a) $Ra=5\times10^7$, (b) $Ra=5\times10^8$ and (c) $Ra=2\times10^9$. Parameters are $\Gamma=1.5$, $E=10^{-6}$ and $Pr=1$.}
\label{fig:wall1}
\end{center}
\end{figure}

\begin{figure}
\begin{center}
\includegraphics[height=0.27\textwidth]{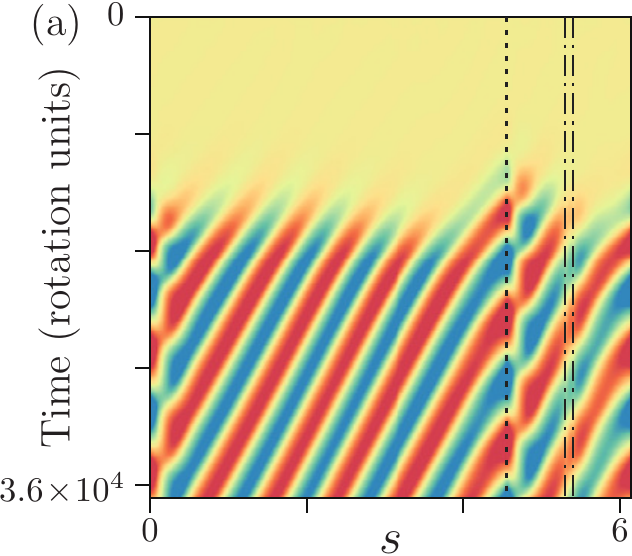}\hfill\includegraphics[height=0.27\textwidth]{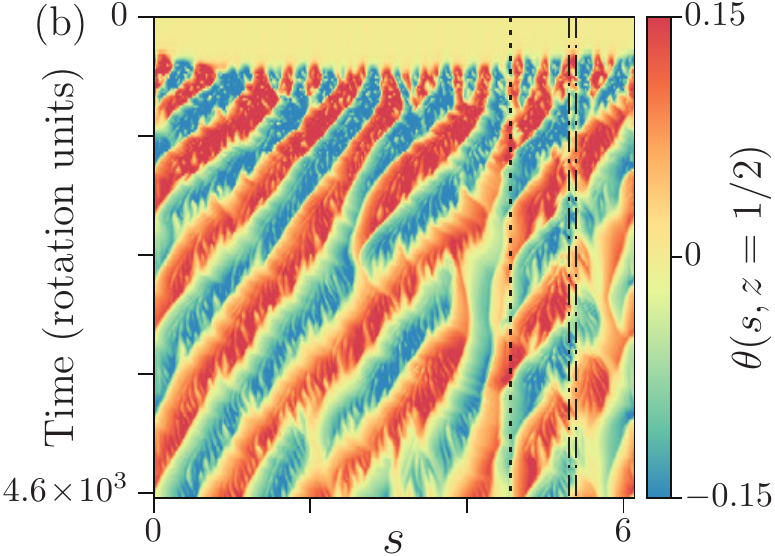}\hfill\includegraphics[height=0.265\textwidth]{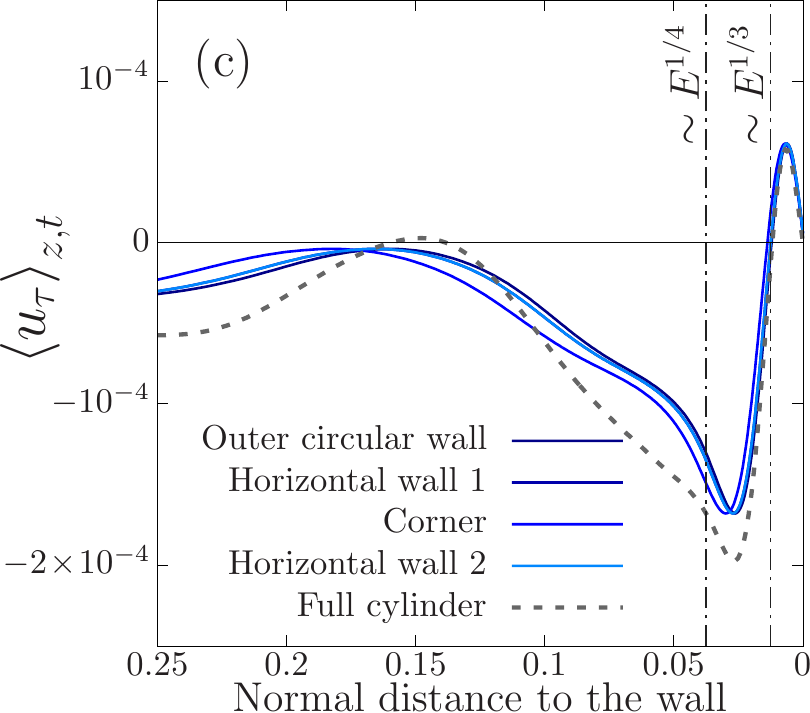}
\caption{Spatio-temporal plots showing the temperature fluctuation $\theta$ at $z=0.5$ and a fixed distance $\delta=10^{-2}$ from the boundary at (a) $Ra=5\times10^7$ and (b) $Ra=5\times10^8$. The horizontal axis represents the arclength $s$ along the boundary while the vertical axis is time (in rotation units). The dotted lines indicate the positions of the four corners of the barrier. Parameters are $\Gamma=1.5$, $E=10^{-6}$ and $Pr=1$. (c) Vertical and temporal average of the velocity component tangential to the boundary $u_{\tau}$ at $Ra=5\times10^8$. We distinguish between the cylindrical boundary and the different faces of the barrier. The results are compared to the case without barrier (see figure~\ref{fig:drift}(c)).}
\label{fig:wall2}
\end{center}
\end{figure}

In this section, we return to the parameters used in section~\ref{sec:main}. However, the cylinder is now truncated and we introduce a rectangular barrier of length $\Gamma/2$ and thickness $\Gamma/20$ along a diameter with one end attached to the circular sidewall (see figure~\ref{fig:wall1}). Like the rest of the sidewall, the barrier is no-slip and insulating. The barrier breaks the rotation invariance of the system and introduces four discontinuities in the curvature of the sidewall. One might therefore expect that this change in the geometry of the domain will result in a drastic change in the nature of the wall state. This is not the case. As shown in figures~\ref{fig:wall1} and \ref{fig:wall2}, for $Ra=5\times10^7$, $Ra=5\times10^8$ and $Ra=2\times10^9$, the wall state persists, in both its weakly nonlinear and fully nonlinear forms. More surprisingly, the drift frequency is the same as in the case without the barrier for all three Rayleigh numbers, as shown in figure~\ref{fig:drift}a. This is even more surprising when one notices that the wavelength of the wall state is comparable with the length of the barrier and larger than its thickness. As seen in figure~\ref{fig:wall1}, the wall mode simply wraps itself around the obstacle, following the same retrograde drift as in the case without a barrier, although its wavelength and frequency along the barrier are both slightly \textit{larger} than around the rest of the cylinder (see the left panels in figure~\ref{fig:wall2}). Moreover, the presence of the corners appears to be responsible for increased shedding of cyclonic disturbances. Despite this the zonal flow discussed earlier is barely modified by the presence of the barrier (see figure~\ref{fig:wall2}b), and its average in a direction tangential to the boundary exhibits the same nested cyclonic-anticyclonic structure along all boundaries. Nevertheless, the amplitude of the zonal flow is very slightly reduced, a fact confirmed in figure~\ref{fig:drift}b by a reduction in its volume average, here performed over the left half of the cylinder only. Finally, as in the case without a barrier, the wall mode survives the onset of the bulk mode and remains visible in instantaneous visualizations of the vertical velocity at the mid-plane even in the presence of bulk turbulence (figure~\ref{fig:wall1}c).

\section{Conclusions}

In this paper, we have explored the nonlinear dynamics of convective wall modes in a rapidly rotating right circular cylinder. These modes appear at lower Rayleigh numbers than the classical bulk modes. We adopt much lower values of the Ekman number ($E=10^{-6}-10^{-7}$) than previous studies, thereby broadening the interval of Rayleigh numbers within which wall modes alone are present. This approach allowed us to study the resulting wall states in the strongly nonlinear regime unencumbered by convection in the bulk.
We observed very rich dynamics with increasing Rayleigh number. Close to the linear onset, we recovered the well-known retrograde wall modes, followed by a series of merging events triggered by an Eckhaus-Benjamin-Feir instability as the Rayleigh number increases. For yet larger Rayleigh numbers, we observed a secondary transition associated with a shear instability of the mean zonal flow driven nonlinearly by the travelling wall state. The resulting complex chaotic structure is confined to the vicinity of the wall and survives the onset of the bulk mode at higher Rayleigh numbers and even bulk turbulence that is present at yet higher Rayleigh numbers.

These observations are reminiscent of recent experimental and numerical studies demonstrating the existence of a so-called boundary zonal flow in high-Rayleigh-number convection in a rotating cylinder \citep{kunnen,bodenschatz}. These studies did not connect the presence of this mean flow with the presence of a wall mode. We have shown here that many of the experimental observations can in fact be explained by the nonlinear dynamics of the wall mode itself, which can be isolated by a suitable choice of parameter values.

Finally, and perhaps most importantly, we have demonstrated that the properties of this nonlinear wall state are remarkably robust with respect to even the most drastic change in the shape of the cylindrical container. Specifically, we have seen that even the introduction of a substantial radial barrier has essentially no effect on the properties of the boundary zonal flow. In this sense these states behave like the topologically protected states in two-dimensional chiral systems even though rotating convection is a three-dimensional nonlinear driven dissipative system. Although this connection will be explored in detail in future work, we mention here that the precessing wall modes studied both here and in the experiments have no inviscid counterpart: the wall mode emerges before the bulk mode because viscosity breaks the geostrophic balance within the Stewartson layer along the sidewall. This makes the wall modes quite unlike the inviscid Kelvin mode in shallow-water theory. \cite{goldstein} show clearly that these modes are well described by ordinary Bessel functions that represent an extended solution of the inviscid Poincar\'e equation. This equation does in fact possess a wall-localized solution described by a modified Bessel function, but this solution precesses in a prograde sense and disappears for unstable stratification \citep{friedlander}.

The behaviour of the nonlinear wall states as the Ekman number is reduced to asymptotically small values, a regime relevant for many geophysical flows, is, of course, of great interest but remains to be studied. In particular, it is known that such rapidly rotating convection can sustain large-scale vortices in the bulk \citep{guervilly_hughes_jones_2014,favier2014pof}; their interaction with the wall states and the accompanying zonal flow remains a numerical and experimental challenge.

\section*{Acknowledgments}
To the authors' knowledge, the robustness of wall modes with respect to changes in geometry was mentioned in a talk by R.~E. Ecke some 20 years ago as recalled by one of us (EK). This work was supported by the France-Berkeley Fund at the University of California, Berkeley. The computations were performed using the HPC resources of GENCI-IDRIS (Grant No. A0080407543). Centre de Calcul Intensif d'Aix-Marseille is acknowledged for granting access to its high-performance computing resources.

\section*{Declaration of interests}
The authors report no conflict of interest.

\bibliography{biblio}
\bibliographystyle{jfm}

\end{document}